\documentclass[pre,twocolumn,aps,superscriptaddress,showpacs]{revtex4-1}
\usepackage{graphicx}
\usepackage{amssymb}
\usepackage{mathrsfs}
\usepackage[english]{babel}
\usepackage{amstext}
\usepackage{amsmath}
\usepackage{bm}
\usepackage{xcolor}
\usepackage{subfig}
\DeclareRobustCommand{\vect}[1]{\bm{#1}}
{

}
\graphicspath{{./}{../}}

\begin{document}

\title{Predicting
plasticity of amorphous solids from instantaneous normal
modes}

\author{Ivan Kriuchevskyi}
\affiliation{Department of Physics ``A. Pontremoli'', University of Milan, via Celoria 16, 20133 Milan, Italy}

\author{Timothy W. Sirk}
\affiliation{Polymers Branch, U.S. Army Research Laboratory, Aberdeen Proving Ground, MD, USA}

\author{Alessio Zaccone}
\affiliation{Department of Physics ``A. Pontremoli'', University of Milan, via Celoria 16, 20133 Milan, Italy}
\affiliation{Department of Chemical Engineering and Biotechnology, University of Cambridge, Cambridge
CB3 0AS, U.K.}

\begin{abstract}
We present a mathematical description of amorphous solid deformation and plasticity by extending the concept of instantaneous normal modes (INMs) to deformed systems, which allows us to retain the effect of strain on the vibrational density of states (VDOS).
Starting from the nonaffine lattice dynamics (NALD) description of elasticity and viscoelasticity of glasses, we formulate the linear response theory up to large deformations by considering the strain-dependent tangent modulus at finite values of shear strain. 
The (nonaffine) tangent shear modulus is computed from the vibrational density of states (VDOS) of affinely strained configurations at varying strain values. The affine strain, found analytically on the static (undeformed) snapshot of the glass, leads to configurations that are rich of soft low-energy modes as well as unstable modes (negative eigenvalues) that are otherwise completely ``washed out'' and lost, if one lets the system fully relax after strain. This procedure is consistent with the structure of NALD.
The INM spectrum of deformed states allows for the analytical prediction of the stress-strain curve of a model glass. Good parameter-free quantitative agreement is shown between the prediction and simulations of athermal quasi-static shear of a coarse-grained polymer glass.
\end{abstract}

 \maketitle

\newpage


\section{Introduction}
Explaining the emergence of rigidity across the glass transition ($T_\text{g}$) and the fact that the low-frequency shear modulus $G$ goes from zero in the liquid to a finite value in the glassy state is one of the overarching goals of condensed matter physics, with widespread applications from materials engineering~\cite{Schuh} to the mechanical stability of amorphous biological matter~\cite{Kaminski}.
An important step towards this goal is to develop a mechanistic understanding of how amorphous solids behave under deformation, i.e. of both their elastic and plastic deformation behaviour. 
In particular, an understanding of how plastic deformation leads to material yielding and what kind of microstructures promote the plastic flow is currently missing, let alone the possibility of predicting the plastic behaviour in terms of stress vs strain. 


While many approaches have aimed at identifying the carriers of plasticity, with moderate success so far given the absence of identifiable microstructures in glass, approaches aiming at describing amorphous plasticity in terms of mechanical instabilities are, with no exceptions, heavily based on numerical simulations which hinders the mechanistic understanding. In this paper, we tackle this problem from a different angle. By exploiting recent success in mathematically describing the temperature-induced softening and melting of glasses based on the so-called Instantaneous Normal Mode (INM) spectrum~\cite{Keyes1997,Stratt1995,Starr2019,PNAS2021}, we apply the same strategy to describe the strain-induced analogues of softening and ``melting'', i.e. the plasticity and yielding phenomena~\cite{Schall}.
 
Starting from the seminal work of Squire, Holt and Hoover~\cite{Squire}, it became clear that, in the case of amorphous systems (and even complex non-centrosymmetric crystals), in addition to the affine displacements \footnote{Affine means that the vector distance between two atoms in the deformed solid is given by the vector distance between two atoms prior to deformation left-multiplied by the strain tensor.}, the mechanical properties are defined by the relaxation of atomic positions towards their equilibrium values, which are called non-affine deformations~\cite{Lemaitre2006}. These deformations make the material softer, i.e. the elastic shear modulus decreases with increasing non-affinity. It turns out that $G = G_\text{A}-G_\text{NA}$, where $G_\text{A}$ is the affine or Born modulus and $-G_\text{NA}$ is the softening correction from nonaffine displacements~\cite{ZacconeScossa2011}. 



\section{Theory}
Despite the fact that a formal expression for the nonaffine corrections was written early on~\cite{Squire}, the concept has been used mainly as a tool to calculate elastic constants in computer simulations~\cite{Lutsko}. Only recently the mathematical nonaffine response theory of viscoelasticity of amorphous solids was developed~\cite{Lemaitre2006,ZacconeScossa2011,Milkus}. It was again limited to small deformations and \emph{athermal}, meaning that the system resides at, or very close to, a local minimum of the potential energy (inherent state). Further introduction of the so-called instantaneous normal modes (INMs) made it possible to extend the theory to finite temperatures up to the glass transition temperature $T_\text{g}$ and slightly beyond~\cite{Prediction2018}. The main idea of INMs is that, instead of characterizing the system in the well-equilibrated inherent states, single snapshots of the non-fully equilibrated system are considered and averaging is performed over the snapshots. This procedure, devised long ago in the context of numerical simulations of liquids~\cite{Keyes1997,Stratt1995} (and recently formulated also analytically~\cite{PNAS2021}) allows one to retain crucial information about anharmonicities and saddle-points, which dominate the dynamics of liquids~\cite{Mizuno_SciPost} and glasses~\cite{Mizuno_INMs_Bulkley,Karmakar}.

The extended theory including INMs (also known as Nonaffine Lattice Dynamics or NALD) yields predictions that agree with coarse-grained molecular dynamics (MD) simulations \cite{Prediction2018,Ivan2020} and with atomistic simulations \cite{ElderACS}, quite well without adjustable parameters. In order to compute the viscoelastic moduli of a model glass of $N$ particles with mass $m$ from the MD configurations with the atoms' positions, one needs to know the vibrational density of states (VDOS) $\rho(\omega)$ and the affine force correlator $\Gamma(\omega)$. This leads to the following expression for the complex viscoelastic modulus $G^*(\Omega)=G'(\Omega)+iG''(\Omega)$ ~\cite{Lemaitre2006,Prediction2018}:
\begin{eqnarray}\label{complexmodulus}
G^*(\Omega)=G_\text{A}-\dfrac{3N}{V} \int_{C}\dfrac{\rho(\omega)  \Gamma(\omega)}{-m\Omega^2 + i \Omega \nu + m\omega^2}d\omega,
\label{eq:GpGpp}
\end{eqnarray}
where $\nu$ is a friction coefficient, the $\Gamma(\omega)$ can be computed if all the eigenvalues and eigenvectors of the Hessian matrix are known, while the VDOS is a modified distribution of eigenvalues. See Ref.\cite{Prediction2018} for details.

A limitation of the theory presented above is that it works only for small deformations. Upon increasing the shear strain ($\gamma$), the amorphous solid can exhibit extensive irreversible plastic deformation. At the moment, there is no way to analytically predict whether a given material state will fail suddenly and catastrophically (brittle failure) or flow like a liquid (ductile yielding). Moreover, we cannot predict when or where it will fail. For disordered solids, including glassy materials, this fundamental question remains a challenge~\cite{Falk_review,Barrat_review,Arratia}. 
  \\

A useful theoretical framework to analyze elementary plastic events is the limit of temperature $T = 0$. To this end, many computational studies on amorphous solids have been performed with the athermal quasi-static (AQS) protocol \cite{Maloney}:  a glass sample initially quenched down to zero temperature is deformed by a quasi-static shear procedure consisting of the (nonaffine) relaxation of the system after each strain step. While this protocol still cannot accurately reproduce the elastic and plastic stress-strain response of real materials due to the missing entropic contributions~\cite{Suter,Rutledge,Sirk2016}, it represents a useful framework for developing a deeper physical understanding of plasticity in amorphous solids~\cite{Manning2020}. As before, the elastic and plastic features of amorphous solids can be understood by analyzing the Hessian matrix~\cite{Manning2020}. In this case, the NALD equation for the shear modulus reads as \cite{Lemaitre2006,ZacconeScossa2011}
\begin{eqnarray}\label{complexmodulus}
G=G_\text{A}-\dfrac{1}{V} \sum_{p}\dfrac{\mathbf{\Xi}^T_p  \cdot \mathbf{\Xi}_p }{ \omega_p^2}.
\label{complexmodulus}
\end{eqnarray}
where ${\omega_p}$ is the $p$-th eigenfrequency, $\mathbf{\Xi}_p$ is the projection of the affine force onto the $p$-th eigenvector of the Hessian, and $V$ is the volume occupied by the system.

Our aim here is to extend the approach to large deformations and hence to predict the stress-strain curve and the yielding point. We propose to construct the INMs spectrum of deformed states (in short, $\gamma$INM) by an instantaneous Affine Transform (AT) from the non-deformed state. This procedure provides a set of deformed configurations $\{r(\gamma^{AT}_i)\}$. Using these configurations in Eq.~\ref{complexmodulus}, we calculate the strain-dependent shear tangent modulus, referred to here as the ``local'' modulus, from which we predict the stress-strain curve. We will subsequently refer to this procedure as $\gamma$NALD. A reconstruction of the whole stress-strain curve of amorphous solids based on modelling the local strain-dependent shear modulus has been presented also in~\cite{Rodney2016}, however their continuum model contained a free parameter given by the size of hypothetical Eshelby inclusions, whereas our prediction is entirely parameter-free and from only microscopic quantities.

We also compared the calculations based on $\gamma$INM  with the calculations obtained with a similar procedure but using, instead of the $\gamma$INM states from AT, the fully relaxed states in the local energy minima or inherent states $\{r(\gamma^{MIN}_i)\}$. We found that this calculation does not predict any softening nor yielding, but just a steady linear elastic regime, because relaxing the configurations at each strain step effectively washes out all the instabilities, the soft modes, and the saddle-points (see below, in Fig. 1) from the VDOS. This is exactly the same as the case of varying temperature at constant density, where the VDOS of fully relaxed configurations is basically $T$-independent~\cite{Prediction2018}. 
Further, using the energy-minimized configurations after each strain step to compute VDOS and shear modulus with Eq. (1) or Eq. (2) would lead to an erroneous ``double counting'' of the nonaffine relaxations. This is because the negative term in Eq. (1) and Eq. (2) already represents the \emph{nonaffine} relaxations from \emph{affine positions} (the $\Xi$ are precisely the force fields that act on the particles in the affine positions~\cite{Lemaitre2006,Maloney}), hence it is consistent that this term is evaluated using the eigenvalues and VDOS of \emph{affinely} deformed configurations.  
Finally, we use the AQS's stress-strain curve as the reference benchmark to test our prediction.


\section{Numerical simulations}
We have used a modified Kremer-Grest model~\cite{Kremer1986} of a coarse-grained polymer system consisting of 100 linear chains of $50$ monomers, where the polymer chain consisted of two masses, chosen as $m_1=1$ and $m_2=3$, placed in an alternating fashion along the chain backbone. 

To test the idea described above, a zero temperature configuration of the solid must first be obtained. All of the above quantities can then be extracted from the coordinate snapshots of the system and knowing the interaction potentials. In brief, the snapshots of the system are obtained using the LAMMPS simulation package\cite{LAMMPS}. After a sufficient number of equilibration steps, the system is slowly quenched below the glass transition temperature $T_\text{g}$, and then the energy minimization is performed. Five replica configurations were constructed, and all results are subsequently averaged over these five glass realizations at $T=0$. 

Each glassy configuration is used as an input for the calculation of the $\gamma$INM. For this we perform an affine transformation (AT) of the initial configuration:
\begin{equation}
\vect{r}(\gamma^{AT}_i)=\vect{\Lambda}\vect{r}(\gamma^{AT}_0=0)
\end{equation}
where $\vect{\Lambda}$ is the simple shear strain transformation matrix (strain tensor), with all diagonal elements equal to $1$, and the only non-zero off-diagonal element  $\vect{\Lambda}_{xy}=\gamma_i$. 
The set of ${\gamma_i}$ values is chosen such that we do not skip any of the significant plastic events. 
For every configuration $\{r(\gamma^{AT}_i)\}$, we calculate the  Hessian matrix $H$ and the affine force field $\mathbf{\Xi}$~\cite{Lemaitre2006}. The Hessian is then diagonalized to obtain the eigefrequencies $\omega_p$, and the eigenvectors needed to compute the $\mathbf{\Xi}_p$ fields projected onto the eigenvectors that enter Eq. (2).

\section{Results}
\subsection{Vibrational density of states under strain}
\begin{figure}
\centering
  \includegraphics[width=.9\linewidth]{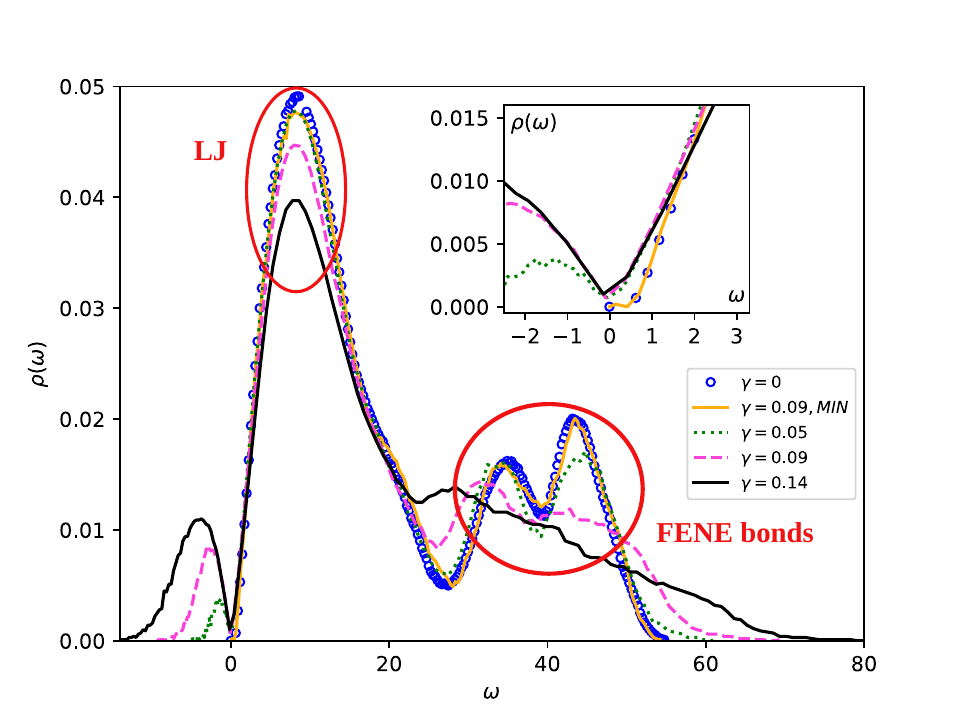}
\caption{VDOS $\rho(\omega)$ for different affine strains $\gamma_i$, showing the INMs spectrum for deformed glasses. Inset gives a closer look at the low-$\omega$ region, where we see the increase of the number of low-energy modes with increase of $\gamma$.
The VDOS for the fully relaxed configuration at $\gamma =0.1$ from the AQS simulation is also shown, and it basically coincides with the VDOS at $\gamma=0$ because the energy minimization at each strain step effectively washes out all the soft low-energy modes and the unstable modes.}
\label{fig:y_VDOS}
\end{figure}

We start by looking at the VDOS of both sets of configurations, $\{r(\gamma^{AT}_i)\}$ and $\{r(\gamma^{MIN}_i)\}$ (Fig. \ref{fig:y_VDOS}). The VDOS for a Kremer-Grest model of polymeric amorphous solid at low temperatures consists of two prominent features: a large peak associated with Lennard-Jones (LJ) interactions between beads and a higher frequency band dominated by FENE bonds vibrations~\cite{length2018}. Also, in the $\{r(\gamma^{AT}_i)\}$ configurations, the diagonalization of the Hessian $H$ produces negative eigenvalues and thus imaginary frequencies. The conventional way of depicting these imaginary frequencies is to show their absolute values on the negative part of the frequency axis as discussed many times in the literature~\cite{Keyes1994,Stratt1995,Keyes1997}.

As shown in Fig. 1, the VDOS of the minimized states from AQS, $\{r(\gamma^{MIN}_i)\}$, does not show signatures of the deformation, similar to what happens as a function of temperature, where the VDOS of well equilibrated systems barely changes with $T$. In contrast, the VDOS of affinely strained (i.e. not fully relaxed) states changes significantly, in the same way as the INMs are traditionally extracted from MD configurations that are not fully relaxed~\cite{Keyes1997,Stratt1995,Starr2019}. Moreover, in the $\gamma$INM procedure, the increase of $\gamma$ produces a similar effect on the VDOS, i.e.  proliferation of soft low-frequency modes, exactly as for the increase of $T$ on the VDOS of liquids and glasses in standard MD simulations at zero strain (for the latter effect see \cite{Prediction2018,Starr2019} and references therein). In particular:
(i) the population of low-energy and saddle-point unstable (negative eigenvalue) modes increases significantly with increasing strain;
(ii) the increase of $\gamma$ causes the  LJ peak and the FENE-bond peaks to decrease and shift to lower frequencies, while a tail of very high-frequency modes emerges at the end of the spectrum and the Debye frequency $\omega_{D}$ is shifted to higher frequencies.\footnote{The increase of $\omega_{D}$ with the affine strain can be explained with the fact that, under a simple shear, bonds in the compression sectors of the solid angle are subjected to compression. It is well known that, in solids under pressure, the frequency of optical-like phonons increases with increasing pressure~\cite{Kunc}, which is important for superconductivity~\cite{Setty}.}

The proliferation of low-energy modes (with positive eigenvalues) directly explains the softening of the material upon increasing $\gamma$, similarly to what happens upon increasing $T$ at vanishing strain as shown in \cite{Prediction2018}.\\

\subsection{Predicting shear modulus and stress-strain curve}

\begin{figure}
  \centering
  \hspace{0.1\textwidth}%
  \centering
  \includegraphics[width=.9\linewidth]{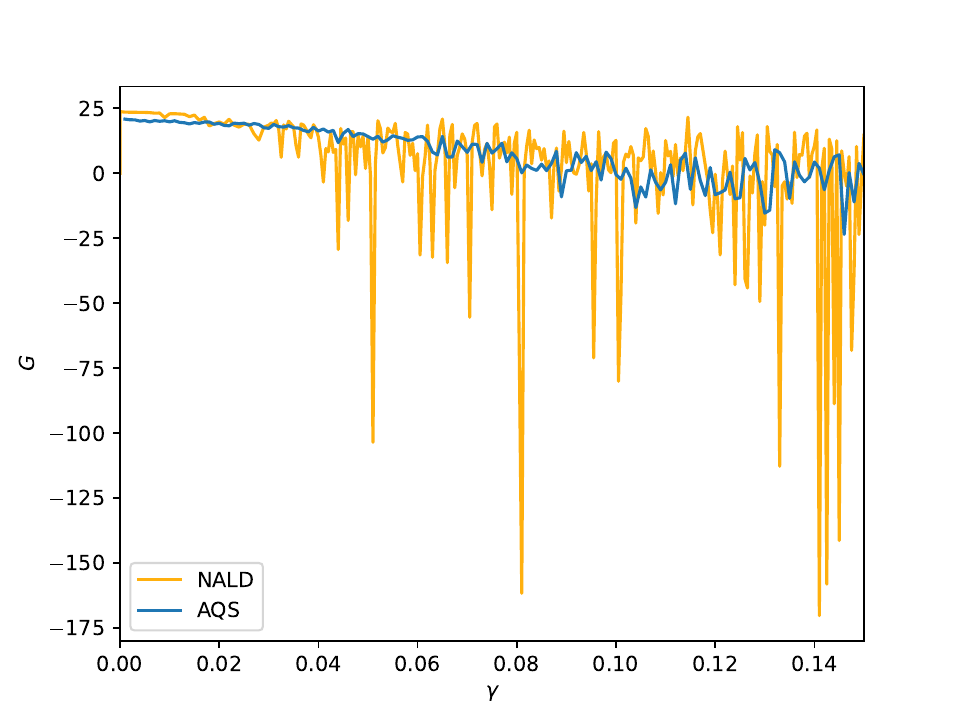}
\caption{Strain-dependent shear modulus. Yellow(light gray) - calculated with Eq.~\ref{complexmodulus} using the $\gamma$INM spectrum as input excluding the lowest negative eigenvalue. Blue(dark gray) - calculated as the slope of AQS stress-strain curve.}
\label{fig:G_y}


\hspace{0.1\textwidth}%
  \centering
  \includegraphics[width=.9\linewidth]{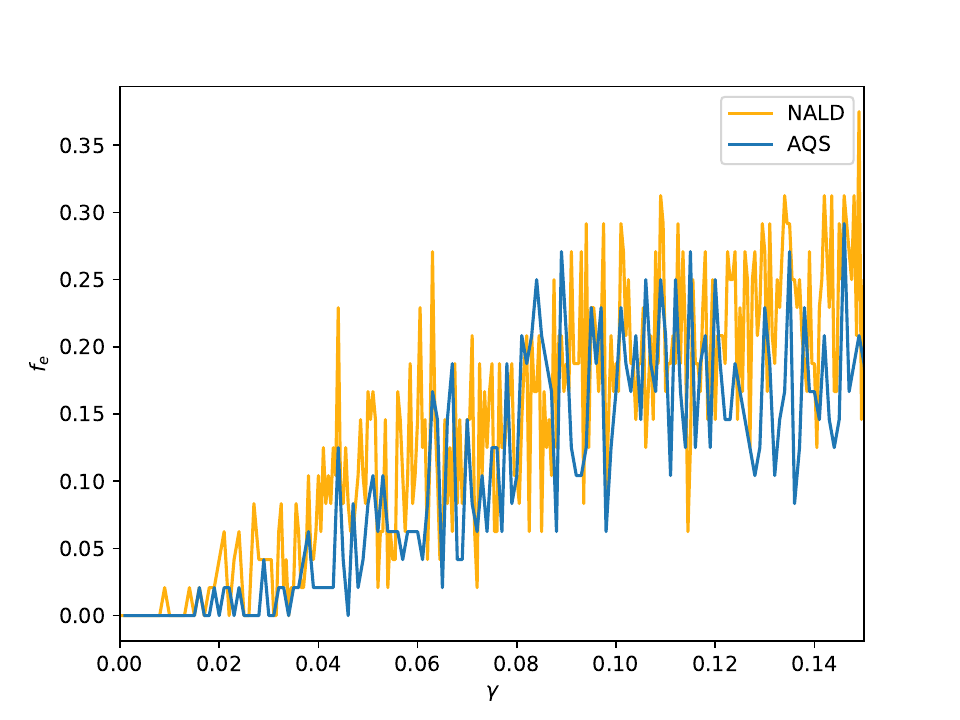}
\caption{The frequency of the plastic events $f_e$, calculated as the average appearance of the negative shear modulus averaged over 50 replicas. }
\label{fig:G_y3}
\hspace{0.1\textwidth}%
\end{figure}
  
\begin{figure}
  \centering
\includegraphics[width=0.9\columnwidth]{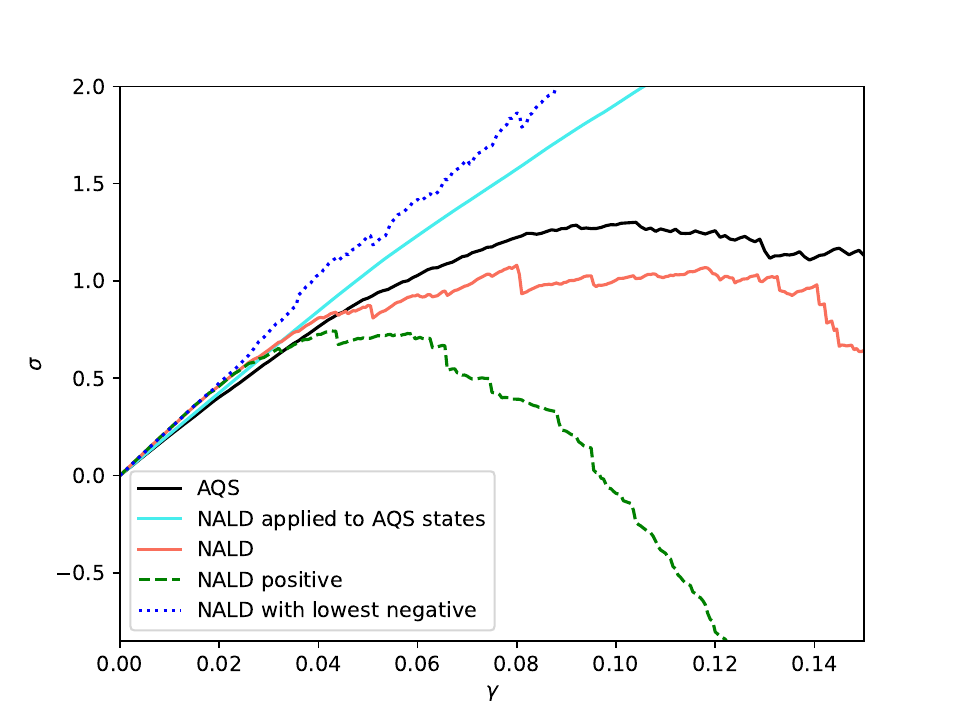}
\caption{Stress vs strain. Stress directly recorded in AQS simulation of deformation (solid black). Stress mathematically predicted from $\gamma$NALD (Eq.~\ref{complexmodulus}) for the set of snapshots $\{r(\gamma^{AT}_i)\}$ of $N=$5k system averaged over 50 replicas. Curves are shown that exclude all negative eigenvalues (dashed green), exclude the lowest negative eigenvalue (red/dark gray), include only the lowest negative eigenvalue (dotted blue) and NALD applied to AQS states (cyan/light gray). }
\label{fig:StressStrain}
\end{figure}

To directly test if the softening predicted by the previous mechanism also occurs in reality, we first semi-analytically calculate the local strain-dependent shear modulus of the $\{r(\gamma^{AT}_i)\}$ configurations using Eq.~\ref{complexmodulus}. Figure \ref{fig:G_y} shows the local shear modulus as a function of the applied shear strain calculated via Eq.~\ref{complexmodulus}, excluding the lowest negative eigenvalue from the sum, and as a slope of the AQS. In both cases we average it over 50 replicas of the polymer glass. As shown in Figure 2, the overall trend of the 'softening' as a manifestation of the strain-induced softening caused by the proliferation of low frequency vibrational modes shown in Fig.\ref{fig:y_VDOS}, similar to the AQS. For some values of strain it drops to  huge negative values, these drops corresponding to a negative slope in the stress-strain curve or, in another words, to mechanical instabilities. We can see that NALD is overestimating the negative drops of the shear modulus. There are two possible reasons for this: 1) the specific AT configurations are not ideal to predict plastic events (but at the moment there are not better alternatives), 2) the Eq.~\ref{complexmodulus} itself is approximate, and possibly must be improved, which is also suggested by the slight discrepancy in the low-$\gamma$ region. Further work is required to clarify this point. 
Note also that these drops of the shear modulus can be called ``plastic events'' only if a single replica is considered. Thus, in Fig.~\ref{fig:G_y3} we show a frequency of the plastic events $f_e$ across 50 replicas, calculated as the average appearance of the negative shear modulus at strain $\gamma$. As shown in the figure, the frequency calculated through $\gamma$NALD is very similar to the one from AQS simulations. 

Using the calculated local shear modulus we then reconstruct the stress via the following algorithm (see also \cite{Rodney2016}):
\begin{equation}
\sigma(\gamma_i)=\sigma(\gamma_{i-1})+ G(\gamma_i)(\gamma_i-\gamma_{i-1}).
\end{equation}

We present the stress-strain curves in Fig.~\ref{fig:StressStrain}. Here, we compare the stress measured as the direct output of the AQS simulation with that predicted from Eq. 4 using the values of $G(\gamma_i)$ computed from the undeformed snapshots. The semi-analytical calculation from Eq. 4 (red/dark gray) gives meaningful results, successfully predicting the deviation from linearity at $\approx 5$\%. Similarly to the vDOS results, using the configurations $\{r(\gamma^{MIN}_i)\}$ from the fully minimized states of the AQS, although it shows a moderate softening, gives no indications at all of the appearance of the plastic events. In a similar way, if we include the lowest negative eigenvalue into our NALD analysis, there are no indications of the softening whatsoever.  \\


\subsection{Analysis of lowest positive and negative eigenvalues}
\begin{figure}
  \centering
\includegraphics[width=0.9\columnwidth]{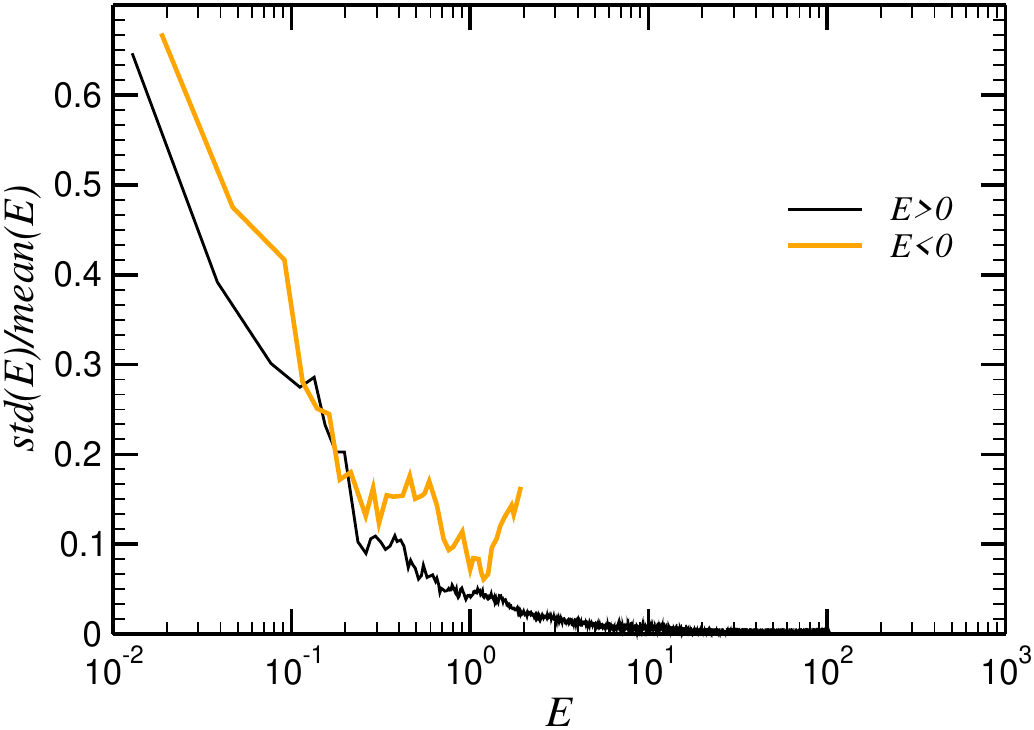}
\caption{Normalized standard deviation for negative and positive eigenvalues. For the negative eigenvalues the absolute value has been taken in the plot. The value of shear strain is fixed at $\gamma=0.03$.}
\label{fig:Estats}
\end{figure}

In order to obtain an agreement between $\gamma$NALD and AQS in the above comparison of stress-strain curve in Fig. 4, the lowest negative eigenvalue was discarded in the $\gamma$NALD calculation. To provide a tentative explanation, we analyzed the statistics of the eigenvalues $E=\omega^2$, both negative $E_{<}$ and positive $E_{>}$, of the Hessian matrix at a fixed strain $\gamma$. The results are shown in Fig. \ref{fig:Estats}. As expected, the standard deviation becomes larger for the modes near $E=0$ (or $\omega=0$). In particular, it is seen that the standard deviation is systematically larger for the negative eigenvalues, i.e. for the INMs, and in particular near $|E|=0$. Upon extrapolating to $|E|=0$, or $\omega=0$, this effect becomes striking, and provides a possible justification for our heuristic elimination of the lowest negative eigenvalues in $\gamma$NALD while retaining the lowest positive one, as argued below.

From the data shown in the figure, we gather that the lowest negative eigenvalue with mean $|E_{<}|=0.0190$ has a non-normalized standard deviation equal to approximately $0.0130$, which is in the same order of magnitude of $|E_{<}|$. 
This, in other words, implies that the lowest negative eigenvalue is statistically less meaningful, which provides a justification for neglecting it in our $\gamma$NALD calculation.

Furthermore, the fact that the standard deviation becomes comparable to the mean, also implies that the lowest negative eigenvalue could be confused for a zero-energy mode. 
This can be explained with the fact that, under a shear $\gamma$ and in a dissipative environment (our polymer glass system is subject to a Langevin thermostat), additional trivial Goldstone-type modes may exist \cite{Minami}, besides the standard three Goldstone trivial modes with $\omega=E=0$ that exist for non-sheared solids due to spontaneously broken translations. Since these three Goldstone modes for non-sheared solids are always discarded in calculations of the dynamics and mechanics, then also any additional Goldstone modes arising from the applied shear field should also be discarded.
This is another possible justification for discarding the lowest negative eigenvalue (and not the lowest positive), i.e. that it may represent a zero Goldstone mode due to shear, as discussed in recent work on the additional Goldstone zero modes in sheared systems, e.g. see \cite{Minami}. Clearly, further work on this issue is required in future investigations.

\section{Conclusion}

In summary, we have presented a microscopic mathematical framework that is able to predict, in a parameter-free way, the nonlinear deformation and plastic flow transition of amorphous solids. 
The approach is based on the nonaffine lattice dynamics (NALD) theory of amorphous solids~\cite{Lemaitre2006,ZacconeScossa2011,Prediction2018,ElderACS} formulated for large strains by extending the concept of Instantaneous Normal Modes (INMs) to deformed glasses. In this procedure, the mechanical relaxations (and avalanches) are effectively taken into account via the imaginary frequencies (unstable modes) contained in the INMs spectrum of the Affinely Transformed (AT) strained configurations, along with the proliferation of low-energy modes upon increasing the strain. These effects can hardly be seen in standard calculations where the energy is minimized after each strain step. Instead, in our approach, by using the affinely transformed strain states (which is the correct input to the nonaffine response calculations), all the information about microscopic relaxation processes is retained.
Using the INMs of the deformed glass as input to the \emph{nonaffine} shear modulus expression in Eq.\eqref{complexmodulus}, it is then possible to semi-analytically reconstruct the stress-strain relation in a parameter-free way via Eq. (4). In comparison with actual AQS simulations of the plastic deformation of a coarse-grained polymer glass, our prediction is able to capture the deviation from the linear elastic regime without the need of performing any simulation of the deformation process, i.e. using only MD snapshots of the undeformed material as input. The methodology is still far from perfect, and there could be better candidates instead of AT configurations to use as the input of the $\gamma$NALD calculation.
In future work, we will investigate systems with more extensive annealing, which have a much more pronounced (sharper) yield point.
It will also be useful to understand how the spatial structure of the lowest eigenmode, allegedly responsible for the softening, is connected to new topological ``defects'' that have been recently identified in the displacement field of deformed glasses, and which can self-organize into a slip system at yielding~\cite{Matteo}. 
Further extensions of the current approach will be useful to elucidate rheological behaviour of soft materials \cite{Mizuno_INMs_Bulkley}.
The $\gamma$NALD approach developed here can be further extended for finite temperatures and for atomistic systems with more complex potentials. \\

\section*{ DATA AVAILABILITY}
The data that supports the findings of this study are available within the article [and its supplementary material].\\

\subsection*{Acknowledgments} 
A.Z. and I.K. acknowledge financial support from US Army Research Laboratory and US Army Research Office through contract nr. W911NF-19-2-0055. 

\bibliography{my}

\begin{thebibliography}{37}%
\makeatletter
\providecommand \@ifxundefined [1]{%
 \@ifx{#1\undefined}
}%
\providecommand \@ifnum [1]{%
 \ifnum #1\expandafter \@firstoftwo
 \else \expandafter \@secondoftwo
 \fi
}%
\providecommand \@ifx [1]{%
 \ifx #1\expandafter \@firstoftwo
 \else \expandafter \@secondoftwo
 \fi
}%
\providecommand \natexlab [1]{#1}%
\providecommand \enquote  [1]{``#1''}%
\providecommand \bibnamefont  [1]{#1}%
\providecommand \bibfnamefont [1]{#1}%
\providecommand \citenamefont [1]{#1}%
\providecommand \href@noop [0]{\@secondoftwo}%
\providecommand \href [0]{\begingroup \@sanitize@url \@href}%
\providecommand \@href[1]{\@@startlink{#1}\@@href}%
\providecommand \@@href[1]{\endgroup#1\@@endlink}%
\providecommand \@sanitize@url [0]{\catcode `\\12\catcode `\$12\catcode
  `\&12\catcode `\#12\catcode `\^12\catcode `\_12\catcode `\%12\relax}%
\providecommand \@@startlink[1]{}%
\providecommand \@@endlink[0]{}%
\providecommand \url  [0]{\begingroup\@sanitize@url \@url }%
\providecommand \@url [1]{\endgroup\@href {#1}{\urlprefix }}%
\providecommand \urlprefix  [0]{URL }%
\providecommand \Eprint [0]{\href }%
\providecommand \doibase [0]{http://dx.doi.org/}%
\providecommand \selectlanguage [0]{\@gobble}%
\providecommand \bibinfo  [0]{\@secondoftwo}%
\providecommand \bibfield  [0]{\@secondoftwo}%
\providecommand \translation [1]{[#1]}%
\providecommand \BibitemOpen [0]{}%
\providecommand \bibitemStop [0]{}%
\providecommand \bibitemNoStop [0]{.\EOS\space}%
\providecommand \EOS [0]{\spacefactor3000\relax}%
\providecommand \BibitemShut  [1]{\csname bibitem#1\endcsname}%
\let\auto@bib@innerbib\@empty
\bibitem [{\citenamefont {Schuh}\ \emph {et~al.}(2007)\citenamefont {Schuh},
  \citenamefont {Hufnagel},\ and\ \citenamefont {Ramamurty}}]{Schuh}%
  \BibitemOpen
  \bibfield  {author} {\bibinfo {author} {\bibfnamefont {C.~A.}\ \bibnamefont
  {Schuh}}, \bibinfo {author} {\bibfnamefont {T.~C.}\ \bibnamefont {Hufnagel}},
  \ and\ \bibinfo {author} {\bibfnamefont {U.}~\bibnamefont {Ramamurty}},\
  }\href {\doibase https://doi.org/10.1016/j.actamat.2007.01.052} {\bibfield
  {journal} {\bibinfo  {journal} {Acta Materialia}\ }\textbf {\bibinfo {volume}
  {55}},\ \bibinfo {pages} {4067} (\bibinfo {year} {2007})}\BibitemShut
  {NoStop}%
\bibitem [{\citenamefont {Chen}\ \emph {et~al.}(2017)\citenamefont {Chen},
  \citenamefont {Young}, \citenamefont {Lu}, \citenamefont {Zaccone},
  \citenamefont {Stroehl}, \citenamefont {Yu}, \citenamefont
  {Kaminski~Schierle},\ and\ \citenamefont {Kaminski}}]{Kaminski}%
  \BibitemOpen
  \bibfield  {author} {\bibinfo {author} {\bibfnamefont {W.}~\bibnamefont
  {Chen}}, \bibinfo {author} {\bibfnamefont {L.~J.}\ \bibnamefont {Young}},
  \bibinfo {author} {\bibfnamefont {M.}~\bibnamefont {Lu}}, \bibinfo {author}
  {\bibfnamefont {A.}~\bibnamefont {Zaccone}}, \bibinfo {author} {\bibfnamefont
  {F.}~\bibnamefont {Stroehl}}, \bibinfo {author} {\bibfnamefont
  {N.}~\bibnamefont {Yu}}, \bibinfo {author} {\bibfnamefont {G.~S.}\
  \bibnamefont {Kaminski~Schierle}}, \ and\ \bibinfo {author} {\bibfnamefont
  {C.~F.}\ \bibnamefont {Kaminski}},\ }\href {\doibase
  10.1021/acs.nanolett.6b03686} {\bibfield  {journal} {\bibinfo  {journal}
  {Nano Letters}\ }\textbf {\bibinfo {volume} {17}},\ \bibinfo {pages} {143}
  (\bibinfo {year} {2017})},\ \bibinfo {note} {pMID: 28073262},\ \Eprint
  {http://arxiv.org/abs/https://doi.org/10.1021/acs.nanolett.6b03686}
  {https://doi.org/10.1021/acs.nanolett.6b03686} \BibitemShut {NoStop}%
\bibitem [{\citenamefont {Keyes}(1997)}]{Keyes1997}%
  \BibitemOpen
  \bibfield  {author} {\bibinfo {author} {\bibfnamefont {T.}~\bibnamefont
  {Keyes}},\ }\href@noop {} {\bibfield  {journal} {\bibinfo  {journal} {J.
  Phys. Chem. A}\ }\textbf {\bibinfo {volume} {101}},\ \bibinfo {pages} {2921}
  (\bibinfo {year} {1997})}\BibitemShut {NoStop}%
\bibitem [{\citenamefont {Stratt}(1995)}]{Stratt1995}%
  \BibitemOpen
  \bibfield  {author} {\bibinfo {author} {\bibfnamefont {R.}~\bibnamefont
  {Stratt}},\ }\href@noop {} {\bibfield  {journal} {\bibinfo  {journal}
  {Macromolecular Theory and Simulations}\ }\textbf {\bibinfo {volume} {28}},\
  \bibinfo {pages} {201–207} (\bibinfo {year} {1995})}\BibitemShut {NoStop}%
\bibitem [{\citenamefont {Zhang}\ \emph {et~al.}(2019)\citenamefont {Zhang},
  \citenamefont {Douglas},\ and\ \citenamefont {Starr}}]{Starr2019}%
  \BibitemOpen
  \bibfield  {author} {\bibinfo {author} {\bibfnamefont {W.}~\bibnamefont
  {Zhang}}, \bibinfo {author} {\bibfnamefont {J.~F.}\ \bibnamefont {Douglas}},
  \ and\ \bibinfo {author} {\bibfnamefont {F.~W.}\ \bibnamefont {Starr}},\
  }\href@noop {} {\bibfield  {journal} {\bibinfo  {journal} {J. Chem. Phys}\
  }\textbf {\bibinfo {volume} {151}},\ \bibinfo {pages} {184904} (\bibinfo
  {year} {2019})}\BibitemShut {NoStop}%
\bibitem [{\citenamefont {Zaccone}\ and\ \citenamefont
  {Baggioli}(2021)}]{PNAS2021}%
  \BibitemOpen
  \bibfield  {author} {\bibinfo {author} {\bibfnamefont {A.}~\bibnamefont
  {Zaccone}}\ and\ \bibinfo {author} {\bibfnamefont {M.}~\bibnamefont
  {Baggioli}},\ }\href {\doibase 10.1073/pnas.2022303118} {\bibfield  {journal}
  {\bibinfo  {journal} {Proceedings of the National Academy of Sciences}\
  }\textbf {\bibinfo {volume} {118}} (\bibinfo {year} {2021}),\
  10.1073/pnas.2022303118},\ \Eprint
  {http://arxiv.org/abs/https://www.pnas.org/content/118/5/e2022303118.full.pdf}
  {https://www.pnas.org/content/118/5/e2022303118.full.pdf} \BibitemShut
  {NoStop}%
\bibitem [{\citenamefont {Denisov}\ \emph {et~al.}(2015)\citenamefont
  {Denisov}, \citenamefont {Dang}, \citenamefont {Struth}, \citenamefont
  {Zaccone}, \citenamefont {Wegdam},\ and\ \citenamefont {Schall}}]{Schall}%
  \BibitemOpen
  \bibfield  {author} {\bibinfo {author} {\bibfnamefont {D.~V.}\ \bibnamefont
  {Denisov}}, \bibinfo {author} {\bibfnamefont {M.~T.}\ \bibnamefont {Dang}},
  \bibinfo {author} {\bibfnamefont {B.}~\bibnamefont {Struth}}, \bibinfo
  {author} {\bibfnamefont {A.}~\bibnamefont {Zaccone}}, \bibinfo {author}
  {\bibfnamefont {G.~H.}\ \bibnamefont {Wegdam}}, \ and\ \bibinfo {author}
  {\bibfnamefont {P.}~\bibnamefont {Schall}},\ }\href@noop {} {\bibfield
  {journal} {\bibinfo  {journal} {Scientific Reports}\ }\textbf {\bibinfo
  {volume} {5}},\ \bibinfo {pages} {14359} (\bibinfo {year}
  {2015})}\BibitemShut {NoStop}%
\bibitem [{\citenamefont {Squire}\ \emph {et~al.}(1969)\citenamefont {Squire},
  \citenamefont {Holt},\ and\ \citenamefont {Hoover}}]{Squire}%
  \BibitemOpen
  \bibfield  {author} {\bibinfo {author} {\bibfnamefont {D.}~\bibnamefont
  {Squire}}, \bibinfo {author} {\bibfnamefont {A.}~\bibnamefont {Holt}}, \ and\
  \bibinfo {author} {\bibfnamefont {W.}~\bibnamefont {Hoover}},\ }\href
  {\doibase https://doi.org/10.1016/0031-8914(69)90031-7} {\bibfield  {journal}
  {\bibinfo  {journal} {Physica}\ }\textbf {\bibinfo {volume} {42}},\ \bibinfo
  {pages} {388 } (\bibinfo {year} {1969})}\BibitemShut {NoStop}%
\bibitem [{Note1()}]{Note1}%
  \BibitemOpen
  \bibinfo {note} {Affine means that the vector distance between two atoms in
  the deformed solid is given by the vector distance between two atoms prior to
  deformation left-multiplied by the strain tensor.}\BibitemShut {Stop}%
\bibitem [{\citenamefont {Lema{\^\i}tre}\ and\ \citenamefont
  {Maloney}(2016)}]{Lemaitre2006}%
  \BibitemOpen
  \bibfield  {author} {\bibinfo {author} {\bibfnamefont {A.}~\bibnamefont
  {Lema{\^\i}tre}}\ and\ \bibinfo {author} {\bibfnamefont {C.}~\bibnamefont
  {Maloney}},\ }\href@noop {} {\bibfield  {journal} {\bibinfo  {journal}
  {Physical Review E}\ }\textbf {\bibinfo {volume} {93}},\ \bibinfo {pages}
  {023006} (\bibinfo {year} {2016})}\BibitemShut {NoStop}%
\bibitem [{\citenamefont {Zaccone}\ and\ \citenamefont
  {Scossa-Romano}(2011)}]{ZacconeScossa2011}%
  \BibitemOpen
  \bibfield  {author} {\bibinfo {author} {\bibfnamefont {A.}~\bibnamefont
  {Zaccone}}\ and\ \bibinfo {author} {\bibfnamefont {E.}~\bibnamefont
  {Scossa-Romano}},\ }\href {\doibase 10.1103/PhysRevB.83.184205} {\bibfield
  {journal} {\bibinfo  {journal} {Phys. Rev. B}\ }\textbf {\bibinfo {volume}
  {83}},\ \bibinfo {pages} {184205} (\bibinfo {year} {2011})}\BibitemShut
  {NoStop}%
\bibitem [{\citenamefont {Lutsko}(1989)}]{Lutsko}%
  \BibitemOpen
  \bibfield  {author} {\bibinfo {author} {\bibfnamefont {J.~F.}\ \bibnamefont
  {Lutsko}},\ }\href {\doibase 10.1063/1.342716} {\bibfield  {journal}
  {\bibinfo  {journal} {Journal of Applied Physics}\ }\textbf {\bibinfo
  {volume} {65}},\ \bibinfo {pages} {2991} (\bibinfo {year} {1989})},\ \Eprint
  {http://arxiv.org/abs/https://doi.org/10.1063/1.342716}
  {https://doi.org/10.1063/1.342716} \BibitemShut {NoStop}%
\bibitem [{\citenamefont {Milkus}\ and\ \citenamefont
  {Zaccone}(2017)}]{Milkus}%
  \BibitemOpen
  \bibfield  {author} {\bibinfo {author} {\bibfnamefont {R.}~\bibnamefont
  {Milkus}}\ and\ \bibinfo {author} {\bibfnamefont {A.}~\bibnamefont
  {Zaccone}},\ }\href {\doibase 10.1103/PhysRevE.95.023001} {\bibfield
  {journal} {\bibinfo  {journal} {Phys. Rev. E}\ }\textbf {\bibinfo {volume}
  {95}},\ \bibinfo {pages} {023001} (\bibinfo {year} {2017})}\BibitemShut
  {NoStop}%
\bibitem [{\citenamefont {Palyulin}\ \emph {et~al.}(2018)\citenamefont
  {Palyulin}, \citenamefont {Ness}, \citenamefont {Milkus}, \citenamefont
  {Elder}, \citenamefont {Sirk},\ and\ \citenamefont
  {Zaccone}}]{Prediction2018}%
  \BibitemOpen
  \bibfield  {author} {\bibinfo {author} {\bibfnamefont {V.~V.}\ \bibnamefont
  {Palyulin}}, \bibinfo {author} {\bibfnamefont {C.}~\bibnamefont {Ness}},
  \bibinfo {author} {\bibfnamefont {R.}~\bibnamefont {Milkus}}, \bibinfo
  {author} {\bibfnamefont {R.~M.}\ \bibnamefont {Elder}}, \bibinfo {author}
  {\bibfnamefont {T.~W.}\ \bibnamefont {Sirk}}, \ and\ \bibinfo {author}
  {\bibfnamefont {A.}~\bibnamefont {Zaccone}},\ }\href@noop {} {\bibfield
  {journal} {\bibinfo  {journal} {Soft Matter}\ }\textbf {\bibinfo {volume}
  {14}},\ \bibinfo {pages} {8475} (\bibinfo {year} {2018})}\BibitemShut
  {NoStop}%
\bibitem [{\citenamefont {Shimada}\ \emph {et~al.}(2021)\citenamefont
  {Shimada}, \citenamefont {Coslovich}, \citenamefont {Mizuno},\ and\
  \citenamefont {Ikeda}}]{Mizuno_SciPost}%
  \BibitemOpen
  \bibfield  {author} {\bibinfo {author} {\bibfnamefont {M.}~\bibnamefont
  {Shimada}}, \bibinfo {author} {\bibfnamefont {D.}~\bibnamefont {Coslovich}},
  \bibinfo {author} {\bibfnamefont {H.}~\bibnamefont {Mizuno}}, \ and\ \bibinfo
  {author} {\bibfnamefont {A.}~\bibnamefont {Ikeda}},\ }\href {\doibase
  10.21468/SciPostPhys.10.1.001} {\bibfield  {journal} {\bibinfo  {journal}
  {SciPost Phys.}\ }\textbf {\bibinfo {volume} {10}},\ \bibinfo {pages} {1}
  (\bibinfo {year} {2021})}\BibitemShut {NoStop}%
\bibitem [{\citenamefont {Oyama}\ \emph {et~al.}(2021)\citenamefont {Oyama},
  \citenamefont {Mizuno},\ and\ \citenamefont {Ikeda}}]{Mizuno_INMs_Bulkley}%
  \BibitemOpen
  \bibfield  {author} {\bibinfo {author} {\bibfnamefont {N.}~\bibnamefont
  {Oyama}}, \bibinfo {author} {\bibfnamefont {H.}~\bibnamefont {Mizuno}}, \
  and\ \bibinfo {author} {\bibfnamefont {A.}~\bibnamefont {Ikeda}},\ }\href
  {\doibase 10.1103/PhysRevLett.127.108003} {\bibfield  {journal} {\bibinfo
  {journal} {Phys. Rev. Lett.}\ }\textbf {\bibinfo {volume} {127}},\ \bibinfo
  {pages} {108003} (\bibinfo {year} {2021})}\BibitemShut {NoStop}%
\bibitem [{\citenamefont {Krishnan}\ \emph {et~al.}(2021)\citenamefont
  {Krishnan}, \citenamefont {Ramola},\ and\ \citenamefont
  {Karmakar}}]{Karmakar}%
  \BibitemOpen
  \bibfield  {author} {\bibinfo {author} {\bibfnamefont {V.~V.}\ \bibnamefont
  {Krishnan}}, \bibinfo {author} {\bibfnamefont {K.}~\bibnamefont {Ramola}}, \
  and\ \bibinfo {author} {\bibfnamefont {S.}~\bibnamefont {Karmakar}},\
  }\href@noop {} {\enquote {\bibinfo {title} {Universal non-debye low-frequency
  vibrations in sheared amorphous solids},}\ } (\bibinfo {year} {2021}),\
  \Eprint {http://arxiv.org/abs/2104.09181} {arXiv:2104.09181
  [cond-mat.dis-nn]} \BibitemShut {NoStop}%
\bibitem [{\citenamefont {Kriuchevskyi}\ \emph {et~al.}(2020)\citenamefont
  {Kriuchevskyi}, \citenamefont {Palyulin}, \citenamefont {Milkus},
  \citenamefont {Elder}, \citenamefont {Sirk},\ and\ \citenamefont
  {Zaccone}}]{Ivan2020}%
  \BibitemOpen
  \bibfield  {author} {\bibinfo {author} {\bibfnamefont {I.}~\bibnamefont
  {Kriuchevskyi}}, \bibinfo {author} {\bibfnamefont {V.~V.}\ \bibnamefont
  {Palyulin}}, \bibinfo {author} {\bibfnamefont {R.}~\bibnamefont {Milkus}},
  \bibinfo {author} {\bibfnamefont {R.~M.}\ \bibnamefont {Elder}}, \bibinfo
  {author} {\bibfnamefont {T.~W.}\ \bibnamefont {Sirk}}, \ and\ \bibinfo
  {author} {\bibfnamefont {A.}~\bibnamefont {Zaccone}},\ }\href {\doibase
  10.1103/PhysRevB.102.024108} {\bibfield  {journal} {\bibinfo  {journal}
  {Phys. Rev. B}\ }\textbf {\bibinfo {volume} {102}},\ \bibinfo {pages}
  {024108} (\bibinfo {year} {2020})}\BibitemShut {NoStop}%
\bibitem [{\citenamefont {Elder}\ \emph {et~al.}(2019)\citenamefont {Elder},
  \citenamefont {Zaccone},\ and\ \citenamefont {Sirk}}]{ElderACS}%
  \BibitemOpen
  \bibfield  {author} {\bibinfo {author} {\bibfnamefont {R.~M.}\ \bibnamefont
  {Elder}}, \bibinfo {author} {\bibfnamefont {A.}~\bibnamefont {Zaccone}}, \
  and\ \bibinfo {author} {\bibfnamefont {T.~W.}\ \bibnamefont {Sirk}},\ }\href
  {\doibase 10.1021/acsmacrolett.9b00505} {\bibfield  {journal} {\bibinfo
  {journal} {ACS Macro Letters}\ }\textbf {\bibinfo {volume} {8}},\ \bibinfo
  {pages} {1160} (\bibinfo {year} {2019})},\ \Eprint
  {http://arxiv.org/abs/https://doi.org/10.1021/acsmacrolett.9b00505}
  {https://doi.org/10.1021/acsmacrolett.9b00505} \BibitemShut {NoStop}%
\bibitem [{\citenamefont {Falk}\ and\ \citenamefont
  {Langer}(2011)}]{Falk_review}%
  \BibitemOpen
  \bibfield  {author} {\bibinfo {author} {\bibfnamefont {M.~L.}\ \bibnamefont
  {Falk}}\ and\ \bibinfo {author} {\bibfnamefont {J.}~\bibnamefont {Langer}},\
  }\href {\doibase 10.1146/annurev-conmatphys-062910-140452} {\bibfield
  {journal} {\bibinfo  {journal} {Annual Review of Condensed Matter Physics}\
  }\textbf {\bibinfo {volume} {2}},\ \bibinfo {pages} {353} (\bibinfo {year}
  {2011})},\ \Eprint
  {http://arxiv.org/abs/https://doi.org/10.1146/annurev-conmatphys-062910-140452}
  {https://doi.org/10.1146/annurev-conmatphys-062910-140452} \BibitemShut
  {NoStop}%
\bibitem [{\citenamefont {Nicolas}\ \emph {et~al.}(2018)\citenamefont
  {Nicolas}, \citenamefont {Ferrero}, \citenamefont {Martens},\ and\
  \citenamefont {Barrat}}]{Barrat_review}%
  \BibitemOpen
  \bibfield  {author} {\bibinfo {author} {\bibfnamefont {A.}~\bibnamefont
  {Nicolas}}, \bibinfo {author} {\bibfnamefont {E.~E.}\ \bibnamefont
  {Ferrero}}, \bibinfo {author} {\bibfnamefont {K.}~\bibnamefont {Martens}}, \
  and\ \bibinfo {author} {\bibfnamefont {J.-L.}\ \bibnamefont {Barrat}},\
  }\href {\doibase 10.1103/RevModPhys.90.045006} {\bibfield  {journal}
  {\bibinfo  {journal} {Rev. Mod. Phys.}\ }\textbf {\bibinfo {volume} {90}},\
  \bibinfo {pages} {045006} (\bibinfo {year} {2018})}\BibitemShut {NoStop}%
\bibitem [{\citenamefont {Galloway}\ \emph {et~al.}(2020)\citenamefont
  {Galloway}, \citenamefont {Ma}, \citenamefont {Keim}, \citenamefont
  {Jerolmack}, \citenamefont {Yodh},\ and\ \citenamefont {Arratia}}]{Arratia}%
  \BibitemOpen
  \bibfield  {author} {\bibinfo {author} {\bibfnamefont {K.~L.}\ \bibnamefont
  {Galloway}}, \bibinfo {author} {\bibfnamefont {X.}~\bibnamefont {Ma}},
  \bibinfo {author} {\bibfnamefont {N.~C.}\ \bibnamefont {Keim}}, \bibinfo
  {author} {\bibfnamefont {D.~J.}\ \bibnamefont {Jerolmack}}, \bibinfo {author}
  {\bibfnamefont {A.~G.}\ \bibnamefont {Yodh}}, \ and\ \bibinfo {author}
  {\bibfnamefont {P.~E.}\ \bibnamefont {Arratia}},\ }\href {\doibase
  10.1073/pnas.2000698117} {\bibfield  {journal} {\bibinfo  {journal}
  {Proceedings of the National Academy of Sciences}\ }\textbf {\bibinfo
  {volume} {117}},\ \bibinfo {pages} {11887} (\bibinfo {year} {2020})},\
  \Eprint
  {http://arxiv.org/abs/https://www.pnas.org/content/117/22/11887.full.pdf}
  {https://www.pnas.org/content/117/22/11887.full.pdf} \BibitemShut {NoStop}%
\bibitem [{\citenamefont {Maloney}\ and\ \citenamefont
  {Lema\^{\i}tre}(2006)}]{Maloney}%
  \BibitemOpen
  \bibfield  {author} {\bibinfo {author} {\bibfnamefont {C.~E.}\ \bibnamefont
  {Maloney}}\ and\ \bibinfo {author} {\bibfnamefont {A.}~\bibnamefont
  {Lema\^{\i}tre}},\ }\href {\doibase 10.1103/PhysRevE.74.016118} {\bibfield
  {journal} {\bibinfo  {journal} {Phys. Rev. E}\ }\textbf {\bibinfo {volume}
  {74}},\ \bibinfo {pages} {016118} (\bibinfo {year} {2006})}\BibitemShut
  {NoStop}%
\bibitem [{\citenamefont {Theodorou}\ and\ \citenamefont
  {Suter}(1986)}]{Suter}%
  \BibitemOpen
  \bibfield  {author} {\bibinfo {author} {\bibfnamefont {D.~N.}\ \bibnamefont
  {Theodorou}}\ and\ \bibinfo {author} {\bibfnamefont {U.~W.}\ \bibnamefont
  {Suter}},\ }\href {\doibase 10.1021/ma00156a026} {\bibfield  {journal}
  {\bibinfo  {journal} {Macromolecules}\ }\textbf {\bibinfo {volume} {19}},\
  \bibinfo {pages} {379} (\bibinfo {year} {1986})},\ \Eprint
  {http://arxiv.org/abs/https://doi.org/10.1021/ma00156a026}
  {https://doi.org/10.1021/ma00156a026} \BibitemShut {NoStop}%
\bibitem [{\citenamefont {in~'t Veld}\ and\ \citenamefont
  {Rutledge}(2003)}]{Rutledge}%
  \BibitemOpen
  \bibfield  {author} {\bibinfo {author} {\bibfnamefont {P.~J.}\ \bibnamefont
  {in~'t Veld}}\ and\ \bibinfo {author} {\bibfnamefont {G.~C.}\ \bibnamefont
  {Rutledge}},\ }\href {\doibase 10.1021/ma0346658} {\bibfield  {journal}
  {\bibinfo  {journal} {Macromolecules}\ }\textbf {\bibinfo {volume} {36}},\
  \bibinfo {pages} {7358} (\bibinfo {year} {2003})},\ \Eprint
  {http://arxiv.org/abs/https://doi.org/10.1021/ma0346658}
  {https://doi.org/10.1021/ma0346658} \BibitemShut {NoStop}%
\bibitem [{\citenamefont {Sirk}\ \emph {et~al.}(2016)\citenamefont {Sirk},
  \citenamefont {Karim}, \citenamefont {Lenhart}, \citenamefont {Andzelm},\
  and\ \citenamefont {Khare}}]{Sirk2016}%
  \BibitemOpen
  \bibfield  {author} {\bibinfo {author} {\bibfnamefont {T.~W.}\ \bibnamefont
  {Sirk}}, \bibinfo {author} {\bibfnamefont {M.}~\bibnamefont {Karim}},
  \bibinfo {author} {\bibfnamefont {J.~L.}\ \bibnamefont {Lenhart}}, \bibinfo
  {author} {\bibfnamefont {J.~W.}\ \bibnamefont {Andzelm}}, \ and\ \bibinfo
  {author} {\bibfnamefont {R.}~\bibnamefont {Khare}},\ }\href {\doibase
  https://doi.org/10.1016/j.polymer.2016.03.024} {\bibfield  {journal}
  {\bibinfo  {journal} {Polymer}\ }\textbf {\bibinfo {volume} {90}},\ \bibinfo
  {pages} {178 } (\bibinfo {year} {2016})}\BibitemShut {NoStop}%
\bibitem [{\citenamefont {Richard}\ \emph {et~al.}(2020)\citenamefont
  {Richard}, \citenamefont {Ozawa}, \citenamefont {Patinet}, \citenamefont
  {Stanifer}, \citenamefont {Shang}, \citenamefont {Ridout}, \citenamefont
  {Xu}, \citenamefont {Zhang}, \citenamefont {Morse}, \citenamefont {Barrat},
  \citenamefont {Berthier}, \citenamefont {Falk}, \citenamefont {Guan},
  \citenamefont {Liu}, \citenamefont {Martens}, \citenamefont {Sastry},
  \citenamefont {Vandembroucq}, \citenamefont {Lerner},\ and\ \citenamefont
  {Manning}}]{Manning2020}%
  \BibitemOpen
  \bibfield  {author} {\bibinfo {author} {\bibfnamefont {D.}~\bibnamefont
  {Richard}}, \bibinfo {author} {\bibfnamefont {M.}~\bibnamefont {Ozawa}},
  \bibinfo {author} {\bibfnamefont {S.}~\bibnamefont {Patinet}}, \bibinfo
  {author} {\bibfnamefont {E.}~\bibnamefont {Stanifer}}, \bibinfo {author}
  {\bibfnamefont {B.}~\bibnamefont {Shang}}, \bibinfo {author} {\bibfnamefont
  {S.~A.}\ \bibnamefont {Ridout}}, \bibinfo {author} {\bibfnamefont
  {B.}~\bibnamefont {Xu}}, \bibinfo {author} {\bibfnamefont {G.}~\bibnamefont
  {Zhang}}, \bibinfo {author} {\bibfnamefont {P.~K.}\ \bibnamefont {Morse}},
  \bibinfo {author} {\bibfnamefont {J.-L.}\ \bibnamefont {Barrat}}, \bibinfo
  {author} {\bibfnamefont {L.}~\bibnamefont {Berthier}}, \bibinfo {author}
  {\bibfnamefont {M.~L.}\ \bibnamefont {Falk}}, \bibinfo {author}
  {\bibfnamefont {P.}~\bibnamefont {Guan}}, \bibinfo {author} {\bibfnamefont
  {A.~J.}\ \bibnamefont {Liu}}, \bibinfo {author} {\bibfnamefont
  {K.}~\bibnamefont {Martens}}, \bibinfo {author} {\bibfnamefont
  {S.}~\bibnamefont {Sastry}}, \bibinfo {author} {\bibfnamefont
  {D.}~\bibnamefont {Vandembroucq}}, \bibinfo {author} {\bibfnamefont
  {E.}~\bibnamefont {Lerner}}, \ and\ \bibinfo {author} {\bibfnamefont {M.~L.}\
  \bibnamefont {Manning}},\ }\href {\doibase 10.1103/PhysRevMaterials.4.113609}
  {\bibfield  {journal} {\bibinfo  {journal} {Phys. Rev. Materials}\ }\textbf
  {\bibinfo {volume} {4}},\ \bibinfo {pages} {113609} (\bibinfo {year}
  {2020})}\BibitemShut {NoStop}%
\bibitem [{\citenamefont {Albaret}\ \emph {et~al.}(2016)\citenamefont
  {Albaret}, \citenamefont {Tanguy}, \citenamefont {Boioli},\ and\
  \citenamefont {Rodney}}]{Rodney2016}%
  \BibitemOpen
  \bibfield  {author} {\bibinfo {author} {\bibfnamefont {T.}~\bibnamefont
  {Albaret}}, \bibinfo {author} {\bibfnamefont {A.}~\bibnamefont {Tanguy}},
  \bibinfo {author} {\bibfnamefont {F.}~\bibnamefont {Boioli}}, \ and\ \bibinfo
  {author} {\bibfnamefont {D.}~\bibnamefont {Rodney}},\ }\href {\doibase
  10.1103/PhysRevE.93.053002} {\bibfield  {journal} {\bibinfo  {journal} {Phys.
  Rev. E}\ }\textbf {\bibinfo {volume} {93}},\ \bibinfo {pages} {053002}
  (\bibinfo {year} {2016})}\BibitemShut {NoStop}%
\bibitem [{\citenamefont {Kremer}\ and\ \citenamefont
  {Grest}(1986)}]{Kremer1986}%
  \BibitemOpen
  \bibfield  {author} {\bibinfo {author} {\bibfnamefont {K.}~\bibnamefont
  {Kremer}}\ and\ \bibinfo {author} {\bibfnamefont {G.~S.}\ \bibnamefont
  {Grest}},\ }\href@noop {} {\bibfield  {journal} {\bibinfo  {journal}
  {Physical Review A}\ }\textbf {\bibinfo {volume} {33}},\ \bibinfo {pages}
  {3628} (\bibinfo {year} {1986})}\BibitemShut {NoStop}%
\bibitem [{\citenamefont {S}(1995)}]{LAMMPS}%
  \BibitemOpen
  \bibfield  {author} {\bibinfo {author} {\bibfnamefont {P.}~\bibnamefont
  {S}},\ }\href@noop {} {\bibfield  {journal} {\bibinfo  {journal} {J Comp
  Phys}\ } (\bibinfo {year} {1995})},\ \bibinfo {note} {see also:
  http://lammps.sandia.gov}\BibitemShut {NoStop}%
\bibitem [{\citenamefont {Milkus}\ \emph {et~al.}(2018)\citenamefont {Milkus},
  \citenamefont {Ness}, \citenamefont {Palyulin}, \citenamefont {Weber},
  \citenamefont {Lapkin},\ and\ \citenamefont {Zaccone}}]{length2018}%
  \BibitemOpen
  \bibfield  {author} {\bibinfo {author} {\bibfnamefont {R.}~\bibnamefont
  {Milkus}}, \bibinfo {author} {\bibfnamefont {C.}~\bibnamefont {Ness}},
  \bibinfo {author} {\bibfnamefont {V.~V.}\ \bibnamefont {Palyulin}}, \bibinfo
  {author} {\bibfnamefont {J.}~\bibnamefont {Weber}}, \bibinfo {author}
  {\bibfnamefont {A.}~\bibnamefont {Lapkin}}, \ and\ \bibinfo {author}
  {\bibfnamefont {A.}~\bibnamefont {Zaccone}},\ }\href {\doibase
  10.1021/acs.macromol.7b02352} {\bibfield  {journal} {\bibinfo  {journal}
  {Macromolecules}\ }\textbf {\bibinfo {volume} {51}},\ \bibinfo {pages} {1559}
  (\bibinfo {year} {2018})},\ \Eprint
  {http://arxiv.org/abs/https://doi.org/10.1021/acs.macromol.7b02352}
  {https://doi.org/10.1021/acs.macromol.7b02352} \BibitemShut {NoStop}%
\bibitem [{\citenamefont {Keyes}(1994)}]{Keyes1994}%
  \BibitemOpen
  \bibfield  {author} {\bibinfo {author} {\bibfnamefont {T.}~\bibnamefont
  {Keyes}},\ }\href {\doibase 10.1063/1.468407} {\bibfield  {journal} {\bibinfo
   {journal} {The Journal of Chemical Physics}\ }\textbf {\bibinfo {volume}
  {101}},\ \bibinfo {pages} {5081} (\bibinfo {year} {1994})},\ \Eprint
  {http://arxiv.org/abs/https://doi.org/10.1063/1.468407}
  {https://doi.org/10.1063/1.468407} \BibitemShut {NoStop}%
\bibitem [{Note2()}]{Note2}%
  \BibitemOpen
  \bibinfo {note} {The increase of $\omega _{D}$ with the affine strain can be
  explained with the fact that, under a simple shear, bonds in the compression
  sectors of the solid angle are subjected to compression. It is well known
  that, in solids under pressure, the frequency of optical-like phonons
  increases with increasing pressure~\cite {Kunc}, which is important for
  superconductivity~\cite {Setty}.}\BibitemShut {Stop}%
\bibitem [{\citenamefont {Minami}\ \emph {et~al.}(2021)\citenamefont {Minami},
  \citenamefont {Nakano},\ and\ \citenamefont {Hidaka}}]{Minami}%
  \BibitemOpen
  \bibfield  {author} {\bibinfo {author} {\bibfnamefont {Y.}~\bibnamefont
  {Minami}}, \bibinfo {author} {\bibfnamefont {H.}~\bibnamefont {Nakano}}, \
  and\ \bibinfo {author} {\bibfnamefont {Y.}~\bibnamefont {Hidaka}},\ }\href
  {\doibase 10.1103/PhysRevLett.126.141601} {\bibfield  {journal} {\bibinfo
  {journal} {Phys. Rev. Lett.}\ }\textbf {\bibinfo {volume} {126}},\ \bibinfo
  {pages} {141601} (\bibinfo {year} {2021})}\BibitemShut {NoStop}%
\bibitem [{\citenamefont {Baggioli}\ \emph {et~al.}(2021)\citenamefont
  {Baggioli}, \citenamefont {Kriuchevskyi}, \citenamefont {Sirk},\ and\
  \citenamefont {Zaccone}}]{Matteo}%
  \BibitemOpen
  \bibfield  {author} {\bibinfo {author} {\bibfnamefont {M.}~\bibnamefont
  {Baggioli}}, \bibinfo {author} {\bibfnamefont {I.}~\bibnamefont
  {Kriuchevskyi}}, \bibinfo {author} {\bibfnamefont {T.~W.}\ \bibnamefont
  {Sirk}}, \ and\ \bibinfo {author} {\bibfnamefont {A.}~\bibnamefont
  {Zaccone}},\ }\href {\doibase 10.1103/PhysRevLett.127.015501} {\bibfield
  {journal} {\bibinfo  {journal} {Phys. Rev. Lett.}\ }\textbf {\bibinfo
  {volume} {127}},\ \bibinfo {pages} {015501} (\bibinfo {year}
  {2021})}\BibitemShut {NoStop}%
\bibitem [{\citenamefont {Kunc}\ \emph {et~al.}(2003)\citenamefont {Kunc},
  \citenamefont {Loa},\ and\ \citenamefont {Syassen}}]{Kunc}%
  \BibitemOpen
  \bibfield  {author} {\bibinfo {author} {\bibfnamefont {K.}~\bibnamefont
  {Kunc}}, \bibinfo {author} {\bibfnamefont {I.}~\bibnamefont {Loa}}, \ and\
  \bibinfo {author} {\bibfnamefont {K.}~\bibnamefont {Syassen}},\ }\href
  {\doibase 10.1103/PhysRevB.68.094107} {\bibfield  {journal} {\bibinfo
  {journal} {Phys. Rev. B}\ }\textbf {\bibinfo {volume} {68}},\ \bibinfo
  {pages} {094107} (\bibinfo {year} {2003})}\BibitemShut {NoStop}%
\bibitem [{\citenamefont {{Setty}}\ \emph {et~al.}(2020)\citenamefont
  {{Setty}}, \citenamefont {{Baggioli}},\ and\ \citenamefont
  {{Zaccone}}}]{Setty}%
  \BibitemOpen
  \bibfield  {author} {\bibinfo {author} {\bibfnamefont {C.}~\bibnamefont
  {{Setty}}}, \bibinfo {author} {\bibfnamefont {M.}~\bibnamefont {{Baggioli}}},
  \ and\ \bibinfo {author} {\bibfnamefont {A.}~\bibnamefont {{Zaccone}}},\
  }\href@noop {} {\bibfield  {journal} {\bibinfo  {journal} {arXiv e-prints}\
  ,\ \bibinfo {eid} {arXiv:2007.04981}} (\bibinfo {year} {2020})},\ \Eprint
  {http://arxiv.org/abs/2007.04981} {arXiv:2007.04981 [cond-mat.supr-con]}
  \BibitemShut {NoStop}%
\end{thebibliography}%

\end{document}